\documentstyle[spie]{article}

\newcommand{\tablenotemark}[1]{\rlap{$^{\rm #1}$}}
\newcommand{\tablenotetext}[2]{\par\small $^{\rm #1}$#2}

\begin{document}


%
%

\input{epsf}

\title{First results from the UBC etch-alignment mosaic CCD} 

\author{S.C.Chapman, G.S.Burley, G.A.H.Walker \\[12pt]
           UBC Physics and Astronomy,
            Vancouver BC  V6T 1Z4\\[12pt] }
\date{}
\maketitle

\begin{abstract}
First imaging results are obtained with a new CCD mosaic prototype (3K x 3k, 15 $\mu$m pixels). The CCDs are aligned using an etched socket alignment technique.
Three different measurements of the alignment are made using star images, test pattern images, and microscope analysis. The CCDs have an angular misalignement of less than 30~ppm. 
The composite device is flat to within
$\pm 3\ \mu$m, with rows/columns oriented to within 20~ppm.
The use of an existing technology with built in precision reduces many of 
the difficulties and expenses typically encountered with mosaic detector 
construction. A new camera being built for the UBC liquid mirror telescope is also described.

\end{abstract}


\section{INTRODUCTION}
Mosaic CCD cameras are now becoming the wide-field imaging norm in astronomy, spurred by the practical device size fabrication limit for a reasonable yield of about 4K x 4K 15 $\mu$m pixels or less. The scientific prospects for these cameras are being exploited to greater degrees as the cameras and control electronics become simpler to fabricate. Programs that are already benefitting from the increased field of view include stellar population studies, galactic structure work, deep galaxy and star counts, searches for low surface brightness galaxies, and searches for gravitational lensing and microlensing. 
However, inadequately aligned CCDs in a mosaic detector can severely compromise the performance in these applications, especially in areas such as gravitational lensing where the lens potential must be reconstructed from the image points. Photometry on objects that span two CCDs will also be limited in accuracy.
The work involved in deconvolving a large format (30') mosaic image with misaligned CCDs can be daunting, injecting some degree of error into the results of the analysis. 

The largest existing CCD mosaic (8K x 8K pixels) uses a mounting scheme developed by Luppino et al.$^1$. It uses custom packages constructed for each CCD from a low expansion alloy, which are attached to a machined mounting block. Alignment screws provide the micro-adjustments to bring the CCDs into position. 
Other mosaics have also been fabricated using a combination 
of precision machining and delicate assembly steps involving micromanipulation of the detectors under a high power microscope$^{2,4}$, or use of an alignment template$^{3}$.
In the optimal case, the resulting assembled mosaic can have pixels along rows and columns aligned on the scale of one pixel width (15 $\mu$m). In practice, the four element MOCAM detector (Canada-France-Hawaii telescope) fabricated using the Luppino mounting scheme has angular 
misalignments between CCDs of as much as 3 pixels (Greg Fahlman, UBC, 
private communication). 

The precision of alignment achievable using etched alignment sockets in a silicon substrate has been demonstrated previously$^{5,6,7}$, where angular misalignments are on the order of 20 ppm (1 $\mu$m variation over a 5cm substrate) and column registration is to within a tenth of a pixel. In this technique, the varying atomic densities associated with the different crystal planes of silicon are exploited to achieve anisotropic chemical etching. The resulting etched sockets have an extremely precise definition with respect to the CAD designed masks.

This paper presents the first imaging results with a prototype mosaic CCD camera based on such an etched alignment technique. An overview of the technique and construction of the prototype camera are presented in section 2. In section 3, the imaging results are discussed. Section 4 describes three techniques for assessing the alignment precision of the mosaic, and section 5 outlines future developments.

\section{UBC 3K x 3K MOSAIC CAMERA}

\subsection{Details of the alignment technique}
There are many issues involved, both in the lithographic fabrication of the etched substrate and the assembly of the mosaic, which are discussed in  Chapman et al. 1997 and Chapman 1996. Here we provide an overview as a context for the prototype camera.

The silicon wafer is first etched with the desired socket layout, followed by the definition of aluminum bond pads and traces around the sockets. Semiconductor lithography techniques are used throughout to transfer mask patterns to the wafer and create etch stops. The fabrication process is not without its difficulties. Problems with the chemical etching include residue formation, non-linear etch rates and hillocks (small silicon growths), all of which can lead to misaligned or tilted CCDs. An optimized EDP etchant and careful cleaning steps will alleviate most of these deleterious effects$^{5,9,10}$. The large sockets can lead to imprecision with the metallization as they tend to funnel the photoresist used to define the etch stop. This problem can be minimized through adjustments to some of the processing steps.

\begin{figure}
\epsfxsize 0.5\hsize
\begin{center}
\makebox{
\epsfbox[0 1 526 248]{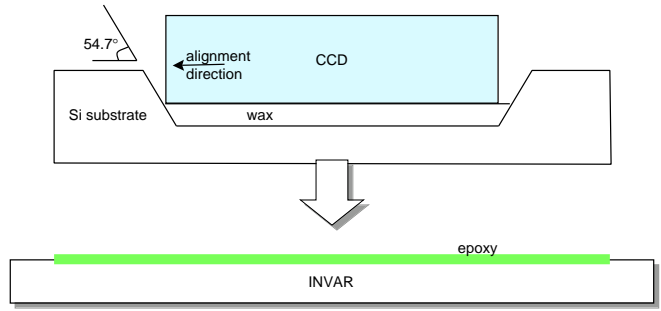}
}
\end{center}
\caption {Cross-section of the mosaic substrate.
Preferential etching along the \{111\} crystal plane yields an 80$\mu$m
deep socket with a 35.3$^\circ$ angled edge.
The substrate is 550$\mu$m thick.
The CCD is butted against the socket edge during the alignment process.}
\label{A}
\end{figure}

Once the substrate has been fabricated and diced from the wafer, the 
mosaic is assembled by first bonding the fragile silicon substrate to an Invar mounting plate with thermally conductive epoxy. 
The substrate is heated and bonding wax applied. The individual CCD detectors are then easily mounted and aligned against two orthogonal reference edges in the sockets, without 
micromanipulation under a high power microscope. Paraffin wax provides a reversible bond, flows well during application and results in a very even layer, which tends to act as a lubricant during the alignment process. 
The detectors are then wire-bonded the to the substrate, or external circuit board. The general alignment concept is shown in Figure~\ref{A}.

\subsection{Description of the camera}
\begin{figure}
\epsfxsize 0.5\hsize
\begin{center}
\makebox{
\epsfbox[48 262 348 546]{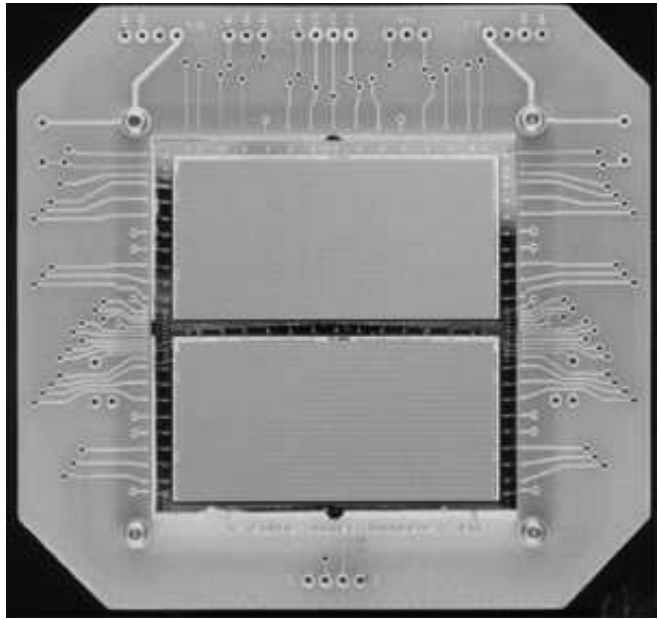}
}
\end{center}
\caption {Photograph of two socket prototype mosaic. The CCDs are wire-
bonded to a circuit board which is fastened to the Invar plate. The CCD
is a
3K $\times$ 1.5K 15$\mu$m pixel detector, with the serial register along
the
long axis.}
\label{C}
\end{figure}

The camera consists of two unthinned Loral 3K x 1.5K, 15$\mu$m pixel CCDs wax bonded to an etched silicon substrate. The flexible silicon substrate is epoxy-bonded to a polished INVAR36 plate to maintain flatness and prevent breakage.
The combination of bond layers, substrate  and Invar plate add a
thermal resistance of approximately 0.1 $^\circ$C/W.
The devices, substrate and mounting plate are bonded evenly over their entire area to form a solid structure. To minimize thermal expansion mismatch stresses, the CTE of the substrate matches that of the detectors (both silicon), and is close to that of the Invar mounting plate (Table 1). At an operating temperature of -90$^\circ$C, we estimate the Invar--silicon thermal expansion mismatch to be about 3$\mu$m over the 5cm substrate length; this has been modeled using finite element analysis (ANSYS 5.3) which shows negligible bowing of the assembly. 
The precision of alignment is detailed in section 4.
The completed prototype is shown in Figure~\ref{C}.

The mosaic detector is mounted to a teflon `spider' and screwed to aluminum brackets in the detector head. Wiring for this prototype is done directly from PCB to the dewar outputs. 
Cooling is achieved with a cold finger attached to a liquid nitrogen 
(LN$_2$) tank. The cold finger quickly cools an aluminum disk, which in turn cools the Invar mounting plate. As the thermal conductivity of the Invar is not high, less robust cooling schemes (such as thermo-electric) may require a more complete Invar surface contact with the cold finger to achieve the -80$^\circ$C operating temperatures. The hold time is roughly 9 hours.
No measurable increase in the outgassing rate was observed compared to the reference empty dewar during thermal cycling$^5$. 

Removal and replacement of CCDs has proven extremely straightforward in practice. Replacement of marginal or faulty devices can be accomplished without directly handling or disturbing the other CCDs. The old wire-bonds are first removed under a microscope. The substrate can then be heated to $\sim50^\circ$ C and the problem CCD replaced.  In practice, it has taken as little as two hours to completely change two CCDs and have the mosaic detector operational again.
Of the seven engineering grade Loral CCDs were made available to us, most had nonfunctional output amplifiers or substrate shorts.
Many of the devices had to be mounted and aligned in the mosaic substrate and then tested in order to find two that worked adequately. The success of the modularity of the technique is assured by the fact that 4 CCDs were replaced in the bottom socket before a working one was found.

\section{FIRST IMAGING RESULTS}
The first images taken with this prototype camera are very promising for the continued and successful use of the etch alignment technique.
Images of various extended objects and star frames were taken in the I-band filter using the UBC 42cm telescope with the mosaic detector mounted at the Cassegrain focus.  The f/13.5 focus provides an image scale of 0.5 arcsec/pixel with a total field of view of 30 arcminutes diameter.
Standard reduction of the images was performed (dark subtraction and flat fielding).

An existing UBC controller built by Ron Johnson was used for initial test purposes. All clock lines were hardwired in parallel on the PC board. The controller only has one channel, thus only the equivalent area of one CCD can be readout at a time. Micro-code has been written for readout of the upper, lower, or half of each CCD for any given exposure. Two frames are shown in Figures~\ref{M13b}, and~\ref{fielda}, of M13 and a random star field.

The images show that cosmetics of the devices are not very good (also see test pattern images, section 4). Numerous dead regions, blocked columns, 
and hot pixels limit the scientific uses. Nonetheless, these CCDs provide a very adequate proof of concept and allow precise measurement of the degree of alignment possible. The mechanical stability of the technique is also verified through repeated use of the camera in lab and on telescope.

\section{THREE MEASUREMENTS OF CCD ALIGNMENT}

In order to test the accuracy of alignment, test images and star field images were taken. Reduction of these images results in viable techniques for measurement of the relative CCD displacement. In addition, measurements under a microscope were used as an external check on the precision. The results of all three techniques are compared in Table 2. The measurements are all in agreement within error estimates. The star fields and optical measurements are considered more reliable than the test patterns for reasons discussed below.

\subsection{Test patterns}

As an initial verification of the CCD angular alignment and assessment of the quality of the CCDs, various test patterns were imaged across the mosaic.
The camera was fitted with a short focal length lens capable of 
imaging a target a few feet away.
As a test of alignment, the data has certain limitations: lens aberrations, non-flat test pattern, and test pattern quality.
The lens gives a vignetted field, only part of which is distortion free. The central region of each image should give images with small aberration. 

\begin{figure}[htbp]
\epsfxsize 0.3\hsize
\begin{center}
\makebox{
\epsfbox[0 0 313 301]{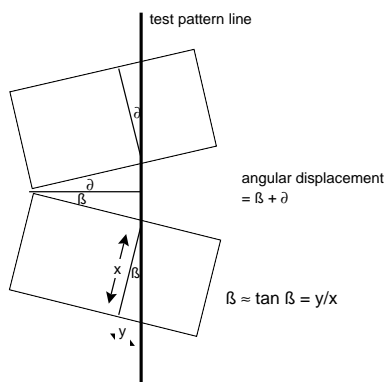}
}
\end{center}
\caption{Geometry for the Two-CCD angular displacement}
\label{triangles}
\end{figure}

\begin{figure}[htbp]
\epsfxsize 0.5\hsize
\begin{center}
\makebox{
\epsfbox[0 0 640 427]{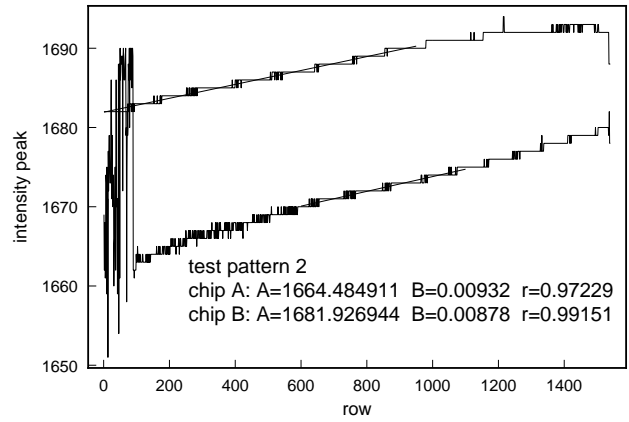}
}
\end{center}
\caption{Reduced data for testpattern 1}
\label{X}
\end{figure}

The data is reduced using the Image Reduction Astronomical Facility (IRAF). The 
test pattern line to be reduced is scanned along each row for a well defined drop-off point in pixel intensity of 20 percent. A linear fit to the central line of the test pattern for each CCD is performed. The coefficients correspond to the formula $y = B x + A$ and $r$ is the correlation coefficient. The two sloped lines on each graph map the column and row of the test pattern on each CCD.
Subtraction of the slope parameters, B, gives the angular misalignment of the two CCDs in the small angle approximation, as shown in figure~\ref{triangles}.
The known distance between CCD imaging areas (2.13mm) is compared with the observed step in the y-intercept of the fit. We find the column registration to be within half a pixel. 
Figure~\ref{X} shows the reduced test pattern data (corresponding to Figures~\ref{XX1},~\ref{XX2}. 

The results are as follows: test1 - 40ppm, test2 - 90ppm, test3 - 50ppm (20ppm corresponds to 1$\mu$m displacement along the 5cm substrate). The spread in values for the 3 tests are more likely an indication of the tests themselves rather than of the actual misalignment. The test patterns are all different and reduction of the data may not provide the same degree of accuracy in all cases. Also, lens aberrations and image flatness may vary between the 3 tests.
We estimate the error for these results to be about $\pm$3 $\mu$m.

\subsection{Star transformations}

\begin{figure}[htbp]
\epsfxsize 0.7\hsize
\begin{center}
\makebox{
\epsfbox[0 0 526 270]{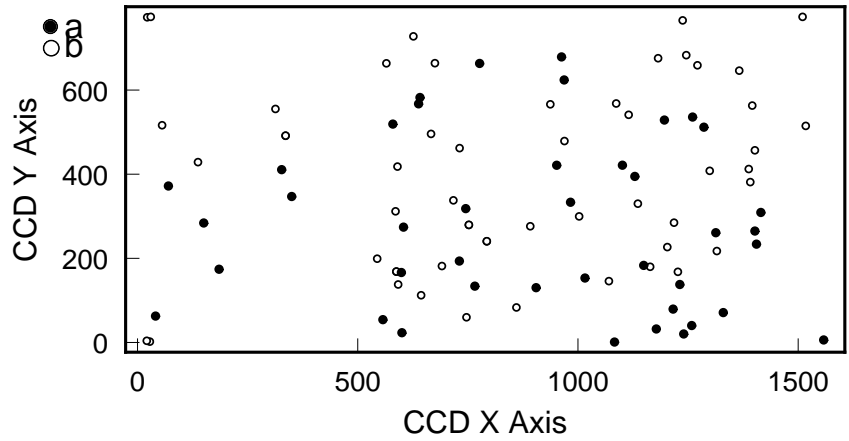}
}
\end{center}
\caption{Star fields from both CCDs superimposed, before transformation}
\label{stars}
\end{figure}

The images of star fields provide a powerful diagnostic of the relative CCD positions. The same telescope and setup are used as described in section 3. The same star field is first imaged on both CCDs by moving the telescope between exposures. By calculating the transformation of one star frame to the other, the angular misalignment of the CCDs can be deduced. The column registration (1.6 $\mu$m) and inter-CCD gap (2.11mm) are found by comparison of features which overlap both CCDs in some frames but not others.
In principle this technique can provide much more information about the relative placement of the CCDs (tip and tilt) and the CCDs themselves (periodic non-uniformities and fringing across the CCD)$^8$. The technique also applies equally well to mosaics of many CCDs. Figure ~\ref{stars} shows the two star fields in the same plane before mapping.

The images were reduced using IRAF and the centroids of all stars were found with DAOPHOT. A table of common sources in both frames is constructed and a transformation is found between the two frames using the routine GEOTRAN in IRAF. Several transformations are found in this way using different numbers of stars in the two frames.
As translation is due to movement of the telescope, the angular rotation of the frames gives the CCD misalignment. The average value is 1.3 $\mu$m over the 5cm substrate (or 26 ppm).
The IRAF routine GEOMAP is used to map one star frame to the other to find an RMS error of 1$\mu$m in the transformation from the mapback accuracy. 

\subsection{Microscope measurements}

A high magnification microscope, with a digitally--metered 
movable eyepiece crosshair accurate to within 0.5$\mu$m, was used to make measurements of the alignment between the detectors 
and the substrate sockets. The etched socket edges were straight and aligned beyond our ability to 
measure them ($\pm$1$\mu$m). 
The angular misalignment of the CCD rows and columns with respect to the 
socket edge was measured to 
be less than 1$\mu$m along the 5cm long axis of each detector (20 parts 
per million). 

The mechanical accuracy of the technique was also verified with a number
of microscope measurements which are detailed 
in Table 3 for the two-element prototype.
Planar (x,y) measurements of the mosaic 
substrate were made using a photographic plate comparator. At low 
magnification, these were repeatable to about a micron. 
Height measurements with 
a calibrated z~stage ($\pm$2$\mu$m accuracy) at the four corners of each device indicate a slight 
tilt to one of the devices along 
the long axis of the CCDs, resulting in a $\pm$3$\mu$m overall flatness 
variation. As this is likely due to 
excessive butting against the angled socket edge,
it may be possible to improve the overall flatness. There was no measurable tip to the CCDs along the short axis of the CCDs.

\section{Future developments}
A 6 meter liquid mirror telescope (LMT) is being constructed near Vancouver BC Canada for a dedicated survey of large scale structure in the universe$^{11}$. 
We are using our mosaic technique to fabricate a 4k x 4k CCD mosaic (four 2kx2k 15$\mu$m pixel CCDs) to operate in time-delay and integrate (TDI) mode$^{12}$.
A widefield corrector lens will be used to correct for distortion of star trails across the CCDs resulting from the high latitude of the telescope. Alternatively, CCDs can easily be angled with respect to each other using the etched socket technique, thereby minimizing the error.
The devices are 2-side buttable, but we are forced to arrange them all with the same orientation for use as a TDI imager. We can therefore only minimize the CCD gap in the N-S orientation. However, the performance of the mosaic is not reduced by more sizable gaps ($\sim$ 1mm) in the E-W direction (readout direction) for TDI readout.

Although all 4 CCDs could be aligned in a common socket, the favored design is to use thin barriers (100 $\mu$m) to maintain modularity of CCDs. Maximum gaps between CCDs of 100 $\mu$m are the same as those proposed for the newest mosaic cameras being fabricated, such as an 8K x 8K pixel mosaic on the 3.5m telescope at the Apache Point Observatory, New Mexico.
CCD spacing of 100 $\mu$m is probably close to the fundamental limitation of the etch technique for CCD packing density. 
The common socket approach would provide excellent alignment if the outer edges of the socket are used as the reference, but at the expense of maximum packing density. Alternatively, if the CCDs are butted against each other, little is gained except in the wax bond and the reference point of the socket edge; the rough CCD edge is not ideal for precision alignment. In both cases, it is very difficult to remove and replace a faulty CCD without disturbing the remaining ones.
 
Other interests have been expressed in using the etch alignment technique to construct mosaic cameras consisting of two 2kx4k pixel CCDs. The technique appears to be adaptable to thinned devices and scaleable to the largest silicon wafer sizes. On an 8" wafer, eight 2k x 4k devices can be aligned. The 12" wafer
awaits 3k x 6k, 15 $\mu$m pixel CCDs.

There are limitations to the technique. Both the precision of angular alignment and the registration of rows/columns depend on the wafer dicing process. Since the device is butted 
against the socket edge, any angular misalignment of the device edge or inconsistency in the imaging area--detector edge distance, will 
show up as misalignment of the 
imaging area.
Based on our measurements of our Loral CCDs, this appears to be a surmountable problem$^5$. 
Height variations of the individual CCD devices will show up as height variations of the composite mosaic. The most realistic solution is to ensure from the manufacturer that the CCDs destined for a mosaic all come from the same thickness wafer.

\section{Summary and Conclusions}
Our successful construction of a prototype camera using the etch alignment  technique provides a sufficient proof of concept to merit further development and use of the technique in future CCD mosaic cameras.
The fabrication process is relatively simple 
and economical using common lithography laboratory equipment. 

Three independent measurements assure the resulting composite device is flat, aligned and mechanically stable. Replacement of faulty CCDs is fast and straightforward.
As far as we are aware, no other existing mosaic 
technique is comparable in terms of alignment achieved.
If the CCDs can be diced accurately enough, there will be accurate registration between pixels and minimal angular displacement between CCDs. 
The overall flatness and alignment of the mosaic are within the tolerances of 
most astronomical observing projects. Software reduction of images should be fast and straightforward. These devices may be suitable for other applications such as medical imaging or remote sensing where large, flat focal plane detectors are critically important.

\section*{Acknowledgments}
With pleasure, we acknowledge the assistance and suggestions of 
Mike Jackson, Paul Hickson, and Ron Johnson.
This research is partially supported by operating grants from the Natural
Sciences and Engineering Research Council of Canada.



\begin{figure}[htbp]
\epsfxsize 0.5\hsize
\begin{center}
\makebox{
\epsfbox[-76 200 689 588]{M13l.ps}
}
\end{center}
\caption{M13, lower CCD }
\label{M13b}
\end{figure}

\begin{figure}[htbp]
\epsfxsize 0.5\hsize
\begin{center}
\makebox{
\epsfbox[-76 203 690 588]{fielda.ps}
}
\end{center}
\caption{star field, upper CCD}
\label{fielda}
\end{figure}

\begin{figure}[htbp]
\epsfxsize 0.5\hsize
\begin{center}
\makebox{
\epsfbox[-76 211 687 588]{test3a.ps}
}
\end{center}
\caption{Testpattern 1 upper CCD }
\label{XX1}
\end{figure}

\begin{figure}[htbp]
\epsfxsize 0.5\hsize
\begin{center}
\makebox{
\epsfbox[-76 237 617 588]{test3b.ps}
}
\end{center}
\caption{Testpattern 1 lower CCD }
\label{XX2}
\end{figure}

\begin{figure}[htbp]
\epsfxsize 0.7\hsize
\begin{center}
\makebox{
\epsfbox[0 0 503 452]{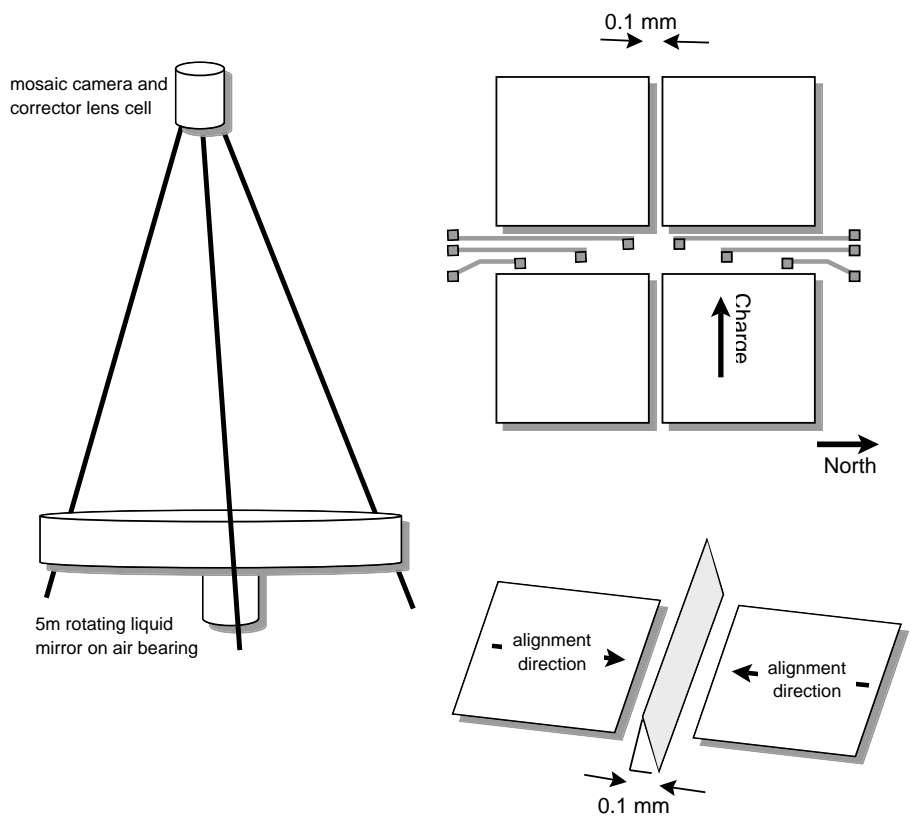}
}
\end{center}
\caption {Clockwise from upper right: Design of the 4kX4k mosaic to be used on the LMT. Detail showing alignment direction on the barriers. Model of the 5m LMT
with mosaic camera at prime focus.}
\label{design}
\end{figure}


\begin{table}[tbp]
\begin{center}
\caption{Material Parameters}
\begin{tabular}{llll}
\\
Material  &  Thickness  &  Conductivity [W$\,m^{-1} K^{-1}$] &  CTE [ppm$\,K^{-1
}$] \\
\hline \hline
Paraffin wax  &  10$\mu$m & 0.25 & 6.7 \\
Silicon & 550$\mu$m & 270 to 170 \tablenotemark{a}& 1.4 to 2.5 \tablenotemark{a}
\\
Invar & 2.5mm & 10.5 & 2.0 to 1.3 \tablenotemark{a}\\
Epoxy & 50$\mu$m & 300 & 4.2\\
\hline
\end{tabular}
\end{center}
 
\label{T1}
\tablenotetext{a}{temperature range: 194 to 273 K}
\end{table}

\begin{table}[h]
\begin{center}
\begin{minipage}{0.8\textwidth}
\renewcommand{\footnoterule}{}
\begin{center}
\caption{Comparison of Three Alignment Measurements}
\begin{tabular*}{\textwidth}{@{\extracolsep{\fill}} lll}
\hline \hline \rule[-8pt]{0cm}{22pt}
Technique  &  Angular alignment \tablenotemark{a}  &  Column Registration  \\
\hline
Test Pattern A \tablenotemark{b} &  2 $\pm$ 3 $\mu$m & 5 $\pm$ 2 $\mu$m \\
B & 4.5 $\mu$m & 8 $\mu$m\\
C & 2.5 $\mu$m & 6 $\mu$m\\
Star Transformations \tablenotemark{c}& 1.3 $\pm$ 1 $\mu$m & $\pm1.6$ $\mu$m \\
Microscope \tablenotemark{d,e}  & 1 $\pm$ 1 $\mu$m & $\pm1.5$ $\mu$m \\
\hline
Average of 3 measures &
1.75 $\mu$m  & 3.13 $\mu$m \\
\hline \hline
\end{tabular*}
\end{center}
\vspace {-1.5ex}
\tablenotetext{a}{Angular deviation measured over the 5cm long axis of each devi
ce}
\tablenotetext{b}{A,B,C refer to 3 different test patterns imaged}
\tablenotetext{c}{Error is RMS for 8 reductions}
\tablenotetext{d}{Alignment measured orthogonal to the serial register direction
}
\tablenotetext{e}{All measurements from detector feature to socket edge}
\end{minipage}
\end{center}
\label{T2}
\end{table}


\begin{table}[h]
\begin{center}
\begin{minipage}{0.8\textwidth}
\renewcommand{\footnoterule}{}
\begin{center}
\caption{Two CCD Mosaic Measurements}
\begin{tabular*}{\textwidth}{@{\extracolsep{\fill}} lll}
\hline \hline \rule[-8pt]{0cm}{22pt}
Parameter  &  Value  &  Units   \\
\hline
Device thickness \tablenotemark{a} \hfill A &  550 $\pm$ 1 & $\mu$m \\
\hfill B & 551 & \mbox{} \\
Substrate thickness & 550 & $\mu$m \\
Socket depth & 80 & $\mu$m \\
Socket flatness & $\pm2$ & $\mu$m \\
Composite device flatness \tablenotemark{b,c} &
$\pm$3  & $\mu$m \\
Device tilt \tablenotemark{c,d} \hfill A & $6\pm2$ & $\mu$m \\
\hfill B & 0 & \mbox{} \\
Device tip \tablenotemark{e} \hfill A,B & $0\pm2$ & $\mu$m \\
\hline \hline
\end{tabular*}
\end{center}
\vspace {-1.5ex}
\tablenotetext{a}{devices are labelled A,B}
\tablenotetext{b}{measured with respect to the substrate}
\tablenotetext{c}{derived from height measurements
at the corners of each device}
\tablenotetext{d}{measured along the serial register direction}
\tablenotetext{e}{measured orthogonal to the serial register direction}
\end{minipage}
\end{center}
\label{T3}
\end{table}

\end{document}